\documentclass[aps,prb,amssymb,showpacs,superscriptaddress,twocolumn,
tightenlines]{revtex4}%
\usepackage{graphicx}
\usepackage{amsmath,bm}
\usepackage[mathscr]{euscript}
\usepackage{dcolumn}

\begin{document}
\title{Theory of Superconductivity in Graphite Intercalation Compounds}
\author{Yasutami Takada}
\thanks{Email: takada@issp.u-tokyo.ac.jp; published in Reference Module 
in Materials Science and Materials Engineering (Saleem Hashmi; editor-in-chief), 
Oxford, Elsevier, 2016; DOI:10.1016/B978-0-12-803581-8.00774-8}
\affiliation{Institute for Solid State Physics, University of Tokyo,
             5-1-5 Kashiwanoha, Kashiwa, Chiba 277-8581, Japan}
\begin{abstract}
On the basis of the model that was successfully applied to KC$_8$, RbC$_8$, and 
CsC$_8$ in 1982, we have calculated the superconducting transition 
temperature $T_c$ for CaC$_6$ and YbC$_6$ to find that the same model 
reproduces the observed $T_c$ in those compounds as well, indicating that 
it is a standard model for superconductivity in the graphite intercalation 
compounds with $T_c$ ranging over three orders of magnitude. Further 
enhancement of $T_c$ well beyond 10 K is also predicted. The present method 
for calculating $T_c$ from first principles is compared with that in the 
density functional theory for superconductors, with paying attention 
to the feature of determining $T_c$ without resort to the concept of the 
Coulomb pseudopotential $\mu^*$.
\end{abstract}



\maketitle

\section{Introduction}
\label{sec:1}

\subsection{Crystal Structure}

For many decades, graphite intercalation compounds (GICs) have been investigated 
from the viewpoints of physics, chemistry, materials science, and engineering 
(or technological) applications~\cite{PhysToday1,PhysToday2,Zabel92,AdvPhys}. Among 
various kinds of GICs, special attention has been paid to the first-stage metal 
compounds, partly because superconductivity is observed mostly in this class of 
GICs, the chemical formula of which is written as $M$C$_x$, where $M$ represents 
either an alkali atom (such as Li, K, Rb, and Cs) or an alkaline-earth atom 
(such as Ca, Sr, Ba, and Yb) and $x$ is either $2$, $6$, or $8$. The crystal 
structure of $M$C$_x$ is shown in Fig.~1(a), in which the metal atom $M$ 
occupies the same spot in the framework of a honeycomb lattice at every $(x/2)$ 
layers of carbon atoms. 
\begin{figure}[htbp]
\begin{center}
\includegraphics[width=8.3cm]{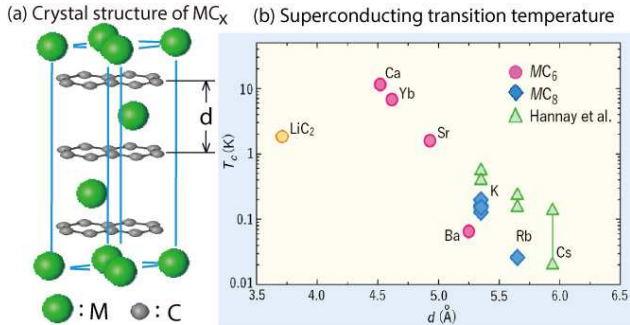}
\end{center}
\caption{(a) Crystal structure of $M$C$_x$（$x=2,6,8$). The case of $x=6$ is 
illustrated here, in which the metal atoms, $M$s, are arranged in a rhombohedral 
structure with the $\alpha \beta \gamma$ stacking sequence, implying that 
$M$ occupies the same spot in the framework of a graphene lattice at every 
three layers (or at the distance of $3d$ with $d$ the distance between the 
adjacent graphite layers). (b) Superconducting transition temperature 
$T_c$ observed in the first-stage alkali- and alkaline-earth-intercalated 
graphites plotted as a function of $d$.}
\label{fig-1}
\end{figure}

\subsection{Superconductivity}

The first discovery of superconductivity in GICs was made in KC$_8$ with the 
superconducting transition temperature $T_c$ of 0.15K in 1965~\cite{Hannay65}. 
In pursuit of higher $T_c$, various GICs were synthesized, mostly working with 
the alkali metals and alkali-metal amalgams as intercalants, from the late 1970s 
to the early 1990s~\cite{Koike78,Kobayashi79,Koike80,Alexander80,Iye,Belash89a,
Dresselhaus89,Belash90}, but only a limited success was achieved at that time; 
the highest attained $T_c$ was around 2-5K in the last century. For example, 
it is 1.9K in LiC$_2$~\cite{Belash89b}. 

A breakthough occurred in 2005 when $T_c$ went up to 11.5K in CaC$_6$~\cite{CaC6,
Emery} (and even to 15.4K under pressures up to 7.5GPa~\cite{Takagi}). In other 
alkaline-earth GICs, the values of $T_c$ are 6.5K, 1.65K, and 0.065K 
for YbC$_6$~\cite{CaC6}, SrC$_6$~\cite{Kim2}, and BaC$_6$~\cite{Heguri}, 
respectively, as indicated in Fig.~1(b). Since then, 
very intensive experimental studies have been made in those and related 
compounds~\cite{Emery,Kim2,Kim1,Kurter}. Theoretical studies have also been 
performed mainly by making state-of-the-art first-principles calculations of 
the electron-phonon coupling constant $\lambda$ to account for the observed 
value of $T_c$ for each individual superconductor~\cite{Mazin,Csanyi,Mauri,
Mauri06,Sanna07}. Those experimental/theoretical works have elucidated that, 
although there are some anisotropic features in the superconducting gap, the 
conventional phonon-driven mechanism to bring about s-wave superconductivity 
applies to those compounds. This picture of superconductivity is confirmed by, 
for example, the observation of the Ca isotope effect with its exponent 
$\alpha=0.50$, the typical Bardeen-Cooper-Schrieffer (BCS) value~\cite{Hinks}. 

\subsection{Central Issues}

In spite of all those efforts and the existence of such a generally accepted 
picture, there remain several very important and fundamental questions: 
\begin{enumerate}
\item 
Can we understand the mechanism of superconductivity in both alkali GICs 
with $T_c$ in the range $0.01-1.0$K and alkaline-earth GICs with $T_c$ typically 
in the range $1-10$K from a unified point of view? In other words, is there 
any standard model for superconductivity in GICs with $T_c$ ranging over three 
orders of magnitude? 

\item
What is the actual reason why $T_c$ is enhanced so abruptly (or by about a 
hundred times) by just substituting K by Ca the atomic mass of which is 
almost the same as that of K? In terms of the standard model, what are the key 
controlling physical parameters to bring about this huge enhancement of $T_c$? 
This change of $T_c$ from KC$_8$ to CaC$_6$ is probably the most important 
issue in exploring superconductivity across the entire family of GICs. 

\item
Is there any possibility to make a further enhancement of $T_c$ in GICs? 
If possible, what is the optimum value of $T_c$ expected in the standard 
model and what kind of atoms should be intercalated to realize the optimum 
$T_c$ in actual GICs? 

\item
What is the physical reason why BaC$_6$ provides so low $T_c$ (=0.065K) 
experimentally, compared with other alkaline-earth GICs and also with 
$T_c=0.23$K in the conventional Eliashberg theory~\cite{Mauri06}? 
This issue may not be important in obtaining high-$T_c$ superconductors, but 
physically it is important enough in comprehensively understanding the 
mechanism of superconductivity in GICs.

\end{enumerate}

In order to provide reliable answers to the above questions, it is indispensable 
to make a first-principles calculation of $T_c$ with sufficient accuracy and 
predictive power. Several years ago, such a calculation was completed by the 
present author and on the basis of the calculation, some interesting predictions 
has been proposed~\cite{Takada09a,Takada09b}. The present chapter not only 
reports some details of this work on the superconducting mechanism in GICs but 
also makes a brief summary of the current status of the theories for 
first-principles calculations of $T_c$. 

\subsection{Organization of This Chapter}

This chapter is organized as follows: In Secs.~2-4, a critical review of the 
theories for quantitative calculations of $T_c$ is given. More specifically, 
we make comments on the theories based on the McMillan's or the Allen-Dynes' 
formula employing the concept of the Coulomb pseudopotential $\mu^*$ in Sec.~2. 
In Sec.~3 we explain the theory on the level of the so-called $G_0W_0$ 
approximation, in which it is a very important merit that $T_c$ can be obtained 
without using $\mu^*$. The same merit can be enjoyed in the density functional 
theory for superconductors, which will be addressed in Sec.~4. In Secs.~5-8, 
a review on superconductivity in GICs is given; starting with a summary of the 
experimental works in Sec.~5, a standard model for considering the mechanism of 
superconductivity in GICs is introduced in Sec.~6. In Sec.~7, the calculated 
results of $T_c$ are given for the alkaline-earth GICs and they are compared 
with the experimental results. The prediction of the optimum $T_c$ is given 
in Sec.~8. Finally in Sec.~9, the conclusion of this chapter is given, together 
with some perspectives on the researches in this and related fields in the 
future. 

\section{First-Principles Calculation of $T_c$}
\label{sec:2}

\subsection{Goal of the Problem}

It would be one of the ultimate goals in the enterprise of condensed matter 
theory to make a reliable prediction of $T_c$ only through the information on 
constituent elements of a superconductor in consideration. A less ambitious yet 
very important goal is to make an accurate evaluation of $T_c$ directly from a 
microscopic (model) Hamiltonian pertinent to the superconductor. If we 
could find the dependence of $T_c$ on the parameters specifying the model 
Hamiltonian, we could obtain a deep insight into the mechanism of 
superconductivity or the competition between the attractive and the repulsive 
interactions between electrons. Accumulation of such information might pave the 
way to the synthesis of a room-temperature superconductor, a big dream in 
materials science. From this perspective, a continuous effort has been made for 
a long time to develop a good theory for first-principles calculations of 
$T_c$, starting with a microscopic Hamiltonian. 

\subsection{McMillan's and Allen-Dynes' Formulas for $T_c$}

In the phonon mechanism, for example, there has been a rather successful 
framework for this purpose, known as the McMillan's formula~\cite{McMillan68} 
or its revised version (the Allen-Dynes' formula)~\cite{Allen75,Allen82,
Carbotte90}, both of which are derived from the Eliashberg theory of 
superconductivity~\cite{Eliashberg60}. In this framework, the task of a 
microscopic calculation of $T_c$ is reduced to the evaluation of the so-called 
Eliashberg function $\alpha^2F(\omega)$ from first principles; the function 
$\alpha^2F(\omega)$ enables us to obtain both the electron-phonon coupling 
constant $\lambda$ and the average phonon energy $\omega_0$, through which we 
can make a first-principles prediction of $T_c$ with an additional introduction 
of a phenomenological parameter $\mu^*$ (the Coulomb 
pseudopotential~\cite{Morel62}) in order to roughly estimate the effect of the 
Coulomb repulsion between electrons on $T_c$. 

At present this framework is usually regarded as the canonical one for making 
a first-principles prediction of $T_c$. In fact, the superconducting mechanism 
of many (so-called weakly-correlated) superconductors is believed to be 
clarified by using this framework, whereby the key phonon modes to bring 
about superconductivity are identified. We can point out that superconductivity 
in MgB$_2$ with $T_c=39$K is a good recent example~\cite{Bohnen01,Kong01,
Choi02a,Choi02b} to illustrate the power of this framework. The case of 
CaC$_6$ has also been investigated along this line of theoretical 
studies~\cite{Mazin,Mauri}. 

\subsection{Coulomb Pseudopotential $\mu^*$}

Nevertheless, this framework is not considered to be very satisfactory, 
primarily because a phenomenological parameter $\mu^*$ is included in the 
theory. Actually, it cannot be regarded as the method of predicting $T_c$ 
in the true sense of the word, if the parameter $\mu^*$ is determined so as to 
reproduce the observed $T_c$. Besides, as long as we employ $\mu^*$ to avoid 
a serious investigation of the effects of the Coulomb repulsion on 
superconductivity, we cannot apply this framework to strongly-correlated 
superconductors. Even in weakly- or moderately-correlated superconductors, 
this framework cannot tell anything about superconductivity originating from 
the Coulomb repulsion via charge and/or spin fluctuations (namely, the 
electronic mechanism including the plasmon mechanism~\cite{Takada78,Takada93a}). 
Furthermore, in this framework, we cannot investigate the competition or the 
coexistence (or even the mutual enhancement due to the quantum-mechanical 
interference effect) between the phonon and the electronic mechanisms. 

The validity of the concept of $\mu^*$ is closely connected with that of the 
Eliashberg theory itself; the theory is valid only when the Fermi energy 
of the superconducting electronic system, $E_{\rm F}$, is very much larger 
than $\omega_0$. Under the condition of $E_{\rm F}\gg \omega_0$, the dynamic 
response time for the phonon-mediated attraction $\omega_0^{-1}$ is much 
slower than that for the Coulomb repulsion $E_{\rm F}^{-1}$, precluding 
any possible interference effects between two interactions, so that physically 
it is very reasonable to separate them. After this separation, the Coulomb part 
(which was not considered to play a positive role in the Cooper-pair formation) 
has been simply treated in terms of $\mu^*$. Thus, for the purpose of searching 
for some positive role of the Coulomb repulsion, the concept of $\mu^*$ is 
irrelevant from the outset. 

\subsection{Vertex Corrections and Dynamic Screening}

Incidentally, in some recently discovered superconductors in the phonon 
mechanism such as the alkali-doped fullerenes with $T_c = 18-38$K~\cite{Hebard91,
Takabayashi09,Gunnarsson97,Takada98}, the condition of $E_{\rm F}\gg \omega_0$ 
is violated, necessitating to include the vertex corrections in calculating 
the phonon-mediated attractive interaction~\cite{Takada93b}. Then, it is 
by no means clear to treat the overall effect of various phonons in terms of 
the sum of the contribution from each phonon, implying that the Eliashberg 
function $\alpha^2F(\omega)$ is not enough to properly describe the attraction 
because of possible interference effects among virtually-excited phonons. As a 
consequence, $\lambda$ will not be simply the sum of $\lambda_i$ the 
contribution from the $i$th phonon, unless $\lambda_i$ is small enough to 
validate the whole calculation in lowest-order perturbation. 

In the case of $E_{\rm F} \approx \omega_0$, another complication occurs in 
treating the screening effect of the conduction electrons. In the usual 
first-principles calculation scheme, the static screening is assumed in 
calculating $\alpha^2F(\omega)$, but it does not reflect the actual screening 
process working during the formation of Cooper pairs. 

\subsection{Ideal Calculation Scheme}

In order to unambiguously solve this problem of screening, we may imagine a 
following ideal calculation scheme for $T_c$: 
In the first step, we calculate the microscopic dynamical electron-electron 
effective interaction $V$ in the whole momentum and energy space. This $V$ is 
assumed to contain both the Coulomb repulsion and the phonon-mediated attraction 
on the same footing. Then in the second step, we obtain $T_c$ directly from this 
$V$ with simultaneously determining the gap function in the whole momentum and 
energy space, reflecting the behavior of $V$. If this scheme were developed, we 
could not only calculate $T_c$ from first principles without resort to $\mu^*$ 
but also correctly discuss the competition, coexistence, and mutual enhancement 
between the phonon and the electronic mechanisms. 

\section{Calculation of $T_c$ in the $G_0W_0$ Approximation}
\label{sec:3}

\subsection{Formulation}

Although it is along the royal road in the project of obtaining a reliable 
method for predicting $T_c$ from first principles, this ideal calculation scheme 
is extremely difficult to achieve in actual situations, because all the 
difficulties in the quantum-mechanical many-body problem are associated with it. 
About three decades ago, the present author, who was a graduate student at that 
time, was struggling with developing such a scheme without perceiving much of 
the difficulties intrinsic to the many-body problem. After a year-long struggle, 
he managed to propose a rather general scheme to evaluate $T_c$ directly from 
$V$ without introducing the concept of $\mu^*$~\cite{Takada78}, though it was 
still at the stage far from the ideal scheme. 

In a broad sense, this scheme may be called an approach from the weak-coupling 
limit, corresponding to the $G_0W_0$ approximation or the one-shot $GW$ 
approximation in the terminology of the present-day first-principles calculation 
community. In the same terminology, by the way, the Eliashberg theory 
corresponds to the $GW$ approximation with respect to the phonon-mediated 
attractive interaction between electrons. 

Let us explain this $G_0W_0$ scheme here~\cite{units}. For simplicity, imagine 
the three-dimensional (3D) electron gas in which an electron is specified by 
momentum ${\bf p}$ and spin $\sigma$. If we write the electron annihilation 
operator by $c_{{\bf p}\sigma}$, we can define the abnormal thermal Green's 
function $F({\bf p},i\omega_p)$ at temperature $T$ by
\begin{eqnarray}
F({\bm p},i\omega_p) = -\int_0^{1/T} \! d\tau \, e^{i\omega_p \tau}
\langle T_{\tau} c_{{\bm p}\uparrow}(\tau)c_{-{\bm p}\downarrow} \rangle,
 \label{eq:F1}
\end{eqnarray}
with $\omega_p$ the fermion Matsubara frequency. At $T=T_c$ where the 
second-order phase transition occurs, this function satisfies the following 
exact gap equation: 
\begin{align}
F({\bm p},i\omega_p) &= -G({\bm p},i\omega_p)G(-{\bm p},-i\omega_p)
\nonumber \\
&\times \!
T_c\! \sum_{\omega_{p'}}\!\sum_{\bm p'}
\tilde{I}({\bm p},{\bm p'};i\omega_p,i\omega_{p'})F({\bm p'},i\omega_{p'}),
\label{eq:F2}
\end{align}
where $G({\bm p},i\omega_p)$ is the normal Green's function 
and $\tilde{I}({\bm p},{\bm p'};i\omega_p,i\omega_{p'})$ is the irreducible 
electron-electron effective interaction. 

Now, in the spirit of the $G_0W_0$ approximation, we replace 
$G({\bm p},i\omega_p)$ by the bare one $G_0({\bm p},i\omega_p)\!\equiv\! 
(i\omega_p\!-\!\varepsilon_{\bm p})^{-1}$ in Eq.~(\ref{eq:F2}), where 
$\varepsilon_{\bm p}(={\bm p}^2/2m^*-\mu)$ is the bare one-electron dispersion 
relation with $m^*$ the band mass and $\mu$ the chemical potential. We will 
also consider the case in which $\tilde{I}({\bm p},{\bm p'};i\omega_p,i\omega_{p'})$ 
is well approximated as a function of only the variables $({\bm p}\!-\!{\bm p'},
i\omega_p\!-\!i\omega_{p'})$ to write 
\begin{eqnarray}
\tilde{I}({\bm p},{\bm p'};i\omega_p,i\omega_{p'})=
V({\bm p}\!-\!{\bm p'},i\omega_p\!-\!i\omega_{p'}),
\label{eq:F2a}
\end{eqnarray}
just like the effective interaction in the random-phase approximation (RPA), 
though we do not intend to confine ourselves to the RPA at this point. By 
substituting Eq.~(\ref{eq:F2a}) into Eq.~(\ref{eq:F2}) and making an 
analytic continuation on the $\omega$ plane to transform $F({\bm p},i\omega_p)$ 
to the retarded function $F^{R}({\bm p},\omega)$ on the real-$\omega$ axis, 
we get a gap equation for $F^{R}({\bm p},\omega)$. Then, by taking the 
imaginary parts in both sides of the gap equation and integrating over the 
$\omega$ variable, we finally obtain an equation depending only on the 
momentum variable ${\bm p}$. Concretely, the equation can be cast into the 
following BCS-type gap equation: 
\begin{eqnarray}
\Delta_{\bm p} = -\sum_{\bm p'} {\Delta_{\bm p'} \over 
2\varepsilon_{\bm p'}} \tanh {\varepsilon_{\bm p'} 
\over 2T_c} \, {\cal K}_{{\bm p},{\bm p'}},
 \label{eq:F3}
\end{eqnarray}
where the gap function $\Delta_{\bm p}$ and the pairing interaction 
${\cal K}_{{\bm p},{\bm p'}}$ are, respectively, defined by 
\begin{eqnarray}
\Delta_{\bm p} \equiv 2|\varepsilon_{\bm p}|
\int_0^{\infty} \! {d\omega \over \pi}\, {\rm Im}\, F^{R}({\bm p},\omega), 
 \label{eq:F4}
\end{eqnarray}
and 
\begin{eqnarray}
{\cal K}_{{\bm p},{\bm p'}} = \int_0^{\infty} \!{2 \over \pi}\,d\Omega\,
{|\varepsilon_{\bm p}|+|\varepsilon_{\bm p'}| 
\over \Omega^2+ 
(|\varepsilon_{\bm p}|+|\varepsilon_{\bm p'}|)^2}\,
V({\bm p}\!-\!{\bm p'},i\Omega)\,.
 \label{eq:F5}
\end{eqnarray}
With use of ${\cal K}_{{\bm p},{\bm p'}}$ thus calculated, we can determine 
$T_c$ as an eigenvalue of Eq.~(\ref{eq:F3}). 

\subsection{Comments on the Formulation}

Five comments are in order on this framework: 
\begin{enumerate}
\item[i)]
Based on Eqs.~(\ref{eq:F3}) and (\ref{eq:F5}), we can obtain $T_c$ directly 
from the microscopic one-electron dispersion relation $\varepsilon_{\bm p}$ 
and the effective electron-electron interaction $V({\bm q},i\Omega)$ with 
no need to separate the phonon-mediated attraction from the Coulomb repulsion. 
\item[ii)]
In spite of the similarity of Eq.~(\ref{eq:F3}) to the BCS gap equation, 
artificial cutoffs involved in constructing the BCS model are avoided in the 
present scheme; natural cutoffs are automatically introduced by the calculation 
of ${\cal K}_{{\bm p},{\bm p'}}$ defined in Eq.~(\ref{eq:F5}). 
\item[iii)]
Except for the spin-singlet pairing, no assumption is made on the 
dependence of the gap function on angular valuables in deriving Eq.~(\ref{eq:F3}), 
so that this gap equation can treat any kind of anisotropy in the gap function, 
indicating that it can be applied to s-wave, d-wave, $\cdots$, and even their 
mixture like (s+d)-wave superconductors. 
\item[iv)]
As can be seen by its definition, the gap function $\Delta_{\bm p}$ 
in Eq.~(\ref{eq:F4}) does not correspond to the physical energy gap except in the 
weak-coupling region. Similary, ${\cal K}_{{\bm p},{\bm p'}}$ is not a physical entity. 
Both quantities are introduced for the mathematical convenience so as to make 
$T_c$ invariant in transforming Eq.~(\ref{eq:F2}) into Eq.~(\ref{eq:F3}). The 
key point here is that we need not solve the full gap equation~(\ref{eq:F2}) 
but much simpler one~(\ref{eq:F3}) in order to obtain $T_c$ in Eq.~(\ref{eq:F2}). 
Of course, if we want to know the physical gap function rather than 
$\Delta_{\bm p}$ to compare with experiment, we need to solve the full gap 
equation, Eq.~(\ref{eq:F2}), with $T_c$ determined by Eq.~(\ref{eq:F3}). 
\item[v)]
Historically, Cohen was the first to evaluate $T_c$ in degenerate semiconductors 
on the level of the $G_0W_0$ approximation~\cite{Cohen64,Cohen69}. 
Unfortunately the pairing interaction is not correctly derived in his 
theory, as explicitly pointed out by the present author~\cite{Takada80} 
who, instead, has succeeded in obtaining the correct pairing 
interaction~\cite{Takada78} by consulting the pertinent work of Kirzhnits 
et al.~\cite{Kirzhnits73}. 
\end{enumerate}

\subsection{Assessment: Application to SrTiO$_3$}

In order to assess the quality of this basic framework of calculating $T_c$ 
from first principles, we have applied it to SrTiO$_3$ and compared the 
results with experiments~\cite{Takada80}. This material is an insulator and 
exhibits ferroelectricity under a uniaxial stress of about 1.6kbar along the 
[100] direction, but it turns into an $n$-type semiconductor by either Nb doping 
or oxygen deficiency, whereby the conduction electrons are introduced in the 
3d band of Ti around the $\Gamma$ point with the band mass of $m^* \approx 
1.8m_e$ ($m_e$: the mass of a free electron). At low temperatures, 
superconductivity appears and the observed $T_c$ shows interesting features; 
$T_c$ depends strongly on the electron concentration $n$ and it is optimized 
with $T_c \approx 0.3$K at $n \approx 10^{20}$cm$^{-3}$. Its dependence on the 
pressure is unsual; $T_c$ decreases rather rapidly with hydrostatic pressures, 
but it increases with the [100] uniaxial stress. 

\begin{figure}[htbp]
\begin{center}
\includegraphics[width=8.5cm]{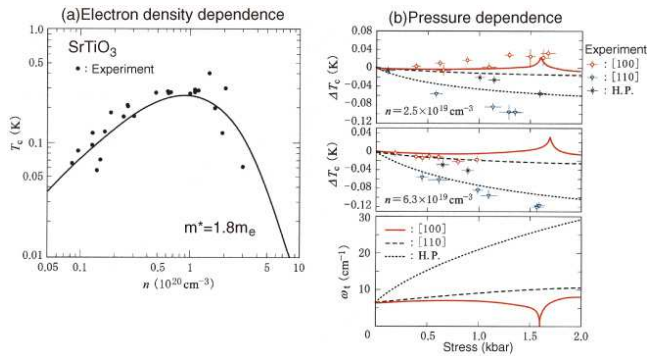}
\end{center}
\caption{(a) Electron density dependence of $T_c$ in SrTiO$_3$ and (b) pressure 
dependence of $T_c$~\cite{Takada80}. Both uniaxial stress along either [100] 
or [110] direction and hydrostatic pressure (H.P.) are considered. The 
deviation of $T_c$, $\Delta T_c$, is plotted as a function of stress in 
units of kbar. For comparison, corresponding experimental results are 
also shown, together with the results of the transverse polar phonon energy 
$\omega_t$ in units of cm$^{-1}$.}
\label{fig-2}
\end{figure}

Taking those situations into account, we have assumed that the superconductivity 
is brought about by the polar-coupling phonons associated with the 
stress-induced ferroelectric phase transition. Then we have calculated the 
effective electron-electron interaction $V({\bm q},i\Omega)$ in the RPA, in 
which the long-range attraction induced by the virtual exchange of polar-coupled 
phonons is included with the long-range Coulomb repulsion on the same footing. 
By substituting this $V({\bm q},i\Omega)$ into Eq.~(\ref{eq:F5}), we have 
obtained $T_c$ directly from a microscopic model and the results of $T_c$ are 
in surprizingly good quantitative agreement with experiment, as shown in 
Fig.~\ref{fig-2}. Recently, a further experimental study on superconductivity 
in this electron-doped SrTiO$_3$ was made to confirm the above results, together 
with the value for the effective mass of $m^*=(1.82\pm 0.05)
m_e$~\cite{Behnia13,Behnia14}. This success indicates that the present basic 
framework including the adoption of the RPA is very useful at least in the 
polar-coupled phonon mechanism. 

\section{Density Functional Theory for Superconductors}
\label{sec:4}

\subsection{Hohenberg-Kohn-Sham Theorem}

Recently much attention has been paid to an extension of the density functional 
theory (DFT) to treat superconductivity, mainly because it provides another 
scheme for first-principles calculations of $T_c$ without resort to $\mu^*$. 
We shall make a very brief review of it in this section. 

It is stated in the basic theorem in DFT that all the physics of an interacting 
electron sysytem is uniquely determined, once its electronic density in the 
ground state $n({\bm r})$ is specified. This Hohenberg-Kohn 
theorem~\cite{Hohenberg64} implies that every physical quantity including the 
exchange-correlation energy $F_{xc}$ may be considered as a unique functional 
of $n({\bm r})$. The density $n({\bm r})$ itself can be determined by solving 
the ground-state electronic density of the corresponding noninteracting 
reference system that is stipulated in terms of the Kohn-Sham (KS) 
equation~\cite{Kohn65}. The core quantity in the KS equation is the 
exchange-correlation potential $V_{xc}({\bm r})$, which is defined as the 
functional derivative of $F_{xc}[n({\bm r})]$ with respect to $n({\bm r})$, 
namely, $V_{xc}({\bm r})=\delta F_{xc}[n]/\delta n({\bm r})$. It must be noted 
that $V_{xc}({\bm r})$ as well as each one-electronic wavefunction at $i$th 
level with its energy eigenvalue $\varepsilon_i$ in the KS equation has no 
physical relevance; they are merely introduced for the mathematical convenience 
so as to obtain the exact $n({\bm r})$ in connecting the noninteracting 
reference system with the real many-electron system. 

The Hohenberg-Kohn theorem can be applied to the ordered ground state as well 
on the understanding that the order parameter itself is regarded as a functional 
of $n({\bm r})$. In providing some approximate functional form for $F_{xc}[n]$, 
however, it would be more convenient to treat the order parameter as an 
additional independent variable. For example, in considering the system with 
some magnetic order, we usually employ the spin-dependent scheme in which the 
fundamental variable is not $n({\bm r})$ but the spin-decomposed density 
$n_{\sigma}({\bm r})$, leading to the spin-polarized exchange-correlation 
energy functional $F_{xc}[n_{\sigma}]$, based on which the spin-dependent KS 
equation is formulated. 

\subsection{Gap Equation}

Similarly, in treating the superconducting state in the framework of DFT, it is 
better to construct the energy functional with employing both $n({\bm r})$ and 
the electron-pair density $\chi({\bm r},{\bm r'}) (\equiv \langle 
\Psi_{\uparrow}({\bm r})\Psi_{\downarrow}({\bm r'})\rangle)$ as basic 
variables, leading to the introduction of the exchange-correlation energy 
functional $F_{xc}[n ({\bm r}),\chi({\bm r},{\bm r'})]$, where 
$\Psi_{\sigma}({\bm r})$ is the electron annihilation 
operator~\cite{Oliveira88,Kurth99}. In accordance with this addition of the 
order parameter as a fundamental variable to DFT, not only the 
exchange-correlation potential $V_{xc}({\bm r})$ but also the 
exchange-correlation pair-potential $\Delta_{xc}({\bm r},{\bm r'})=
-\delta F_{xc}[n,\chi]/\delta \chi^*({\bm r},{\bm r'})$ appear in an extended KS 
equation, which is found to be written in the form of the Bogoliubov-de Gennes 
equation appearing in the usual theory for inhomogeneous 
superconductors~\cite{deGennes66}. Just as is the case with $V_{xc}({\bm r})$, 
$\Delta_{xc}({\bm r},{\bm r'})$ has no direct physical meaning, but in principle, 
if the exact form of $F_{xc}[n,\chi]$ is known, the solution of the extended 
KS equation gives us the exact result for $\chi({\bm r},{\bm r'})$, containing all 
the effects of the Coulomb repulsion including the one usually treated 
phenomenologically through the concept of $\mu^*$. As a result, we can 
determine the exact $T_c$ by the calculation of the highest temperature below 
which a nonzero solution for $\chi({\bm r},{\bm r'})$ can be found. 

In this formulation, we can write the fundamental gap equation to determine 
$T_c$ exactly as
\begin{eqnarray}
\Delta_i = -\sum_{j} {\Delta_j \over 
2\varepsilon_j} \tanh {\varepsilon_j 
\over 2T_c} \, {\cal K}_{ij},
 \label{eq:F6}
\end{eqnarray}
where $\Delta_i$ is the gap function for $i$th KS level. In just 
the same way as its energy eigenvalue $\varepsilon_i$ (which is measured from 
the chemical potential), $\Delta_i$ is not the quantity to be observed 
experimentally but just introduced for the mathematical convenience so as to 
obtain the exact $T_c$ by solving this BCS-type equation, Eq.~(\ref{eq:F6}). 
Similarly, the pair interaction ${\cal K}_{ij}$, defined as the 
second-functional derivative of $F_{xc}[n ,\chi]$ with respect to $\chi^*$ and 
$\chi$, has not any direct physical meaning, either. 

We note here the very impressive fact that the final forms for the two gap 
equations, Eqs.~(\ref{eq:F3}) and (\ref{eq:F6}), are exactly the same, in 
spite of the fact that they are derived from quite different foundations and 
reasoning. We also note that because of this similarity, we may judge that, 
as long as ${\cal K}_{ij}$ is properly chosen, the physics descibed by $\mu^*$ 
is also included in the framework of DFT for superconductors, at least to the 
extent that it is included in the $G_0W_0$ scheme explained in Sec.~3. 

\subsection{Applications}

In 2005, this DFT framework was extended to explicitly taking care of the 
phonon-mediated attractive interaction~\cite{Luders05} and it has been applied 
to many superconductors~\cite{Sanna07,Marques05,Floris05,Profeta06,Sanna06,
Floris07}. In order to perform these calculations for actual superconductors, 
it is necessary to provide a concrete form for $F_{xc}[n ,\chi]$. In the 
judgement of the present author, the presently available form for 
$F_{xc}[n ,\chi]$ contains the information equivalent to that included 
in the Eliashberg theory for the part of the phonon-mediated attraction, 
indicating that no vertex corrections are considered in this treatment, 
while for the part of the Coulomb repulsion, it contains only very crude 
physics; the screening effect is treated in the Thomas-Fermi static-screening 
approximation, which is nothing but the result of the RPA only in the 
static and the long-wavelength limit, forgetting the detailed dynamical 
nature of the screening effect. Mainly for this reason, $T_c$ in the present 
form of $F_{xc}[n ,\chi]$ is not expected to be very accurate, even though 
the calculated results for $T_c$ seem to be in good agreement with expeiment. 

\subsection{Basic Problems}

In relation to the above point, it would be appropriate to give a following 
comment: In the calculations of the normal-state properties in the 
local-density approximation (LDA) and generalized gradient approximation 
(GGA)~\cite{Perdew96} to DFT, we usually anticipate that errors in the 
calculated results are of the order of 1eV and 0.3eV for LDA and GGA, 
respectively. Those errors are much larger than that expected in the 
calculation of quantum chemistry ($\approx 0.05$eV). In DFT for 
superconductors, calculations of $T_c$ (which is of the order of 0.001eV 
in general) are done simultaneously with those for the normal-state 
properties. This implies that the errors anticipated for $T_c$ 
would be very large compared to $T_c$ itself. 

We should also point out that the present form for $F_{xc}[n ,\chi]$ is 
useless to discuss the electronic mechanisms like the plasmon and the 
spin-fluctuation ones, prompting us to improve on the approximate form for 
$F_{xc}[n ,\chi]$. Very recently, a limited improvement on 
$F_{xc}[n ,\chi]$ was made by the inclusion of the contribution from 
plasmons, leading to better agreement with experiment for 
$T_c$~\cite{Arita13,Arita14}. 

Apart from the functional form, there are also several problems in the fundamental 
theory; for example, it is by no means clear whether the second-functional 
derivative of $F_{xc}[n ,\chi]$ is a well-defined quantity or not, in just 
the same way as we have already experienced in the energy-gap 
problem~\cite{Perdew83,Sham83,Sham85} in semiconductors and insulators.

\section{Experiment on Superconductivity in GICs}
\label{sec:5}

From this section, let us get back to the review on superconductivity in GICs. 
As briefly mentioned in Sec.~1, the history of the researches on this issue 
extends more than four decades. In 1965, the first report of superconductivity 
was made for KC$_8$, RbC$_8$, and CsC$_8$~\cite{Hannay65}, in which $T_c$ was 
not reliably determined; it depended very much on samples. Subsequent 
works~\cite{Koike78,Koike80,Kobayashi79,Kobayashi81,Pendrys81,Alexander81,Iye,
Chaiken90} confirmed the occurrence of superconductivity in KC$_8$ with 
$T_c=0.15$K, but superconductivity did not appear in RbC$_8$ and CsC$_8$ down 
to 0.09K and 0.06K, respectively. Later works have found that $T_c$ is actually 
26mK for RbC$_8$~\cite{Zabel92}, but no superconductivity is found in either 
LiC$_6$ or the second- or higher-stage alkali GICs, though the calculation of 
$T_c$ based on the McMillan's formula~\cite{McMillan68} predicted an observable 
value of $T_c$ even for KC$_{24}$~\cite{Kamimura80,Inoshita81}. It seems that 
the usual first-principles calculation of $\alpha^2F(\omega)$ tends to provide 
an unrealistically large contribution from the intralayer high-energy carbon 
oscillations to $\lambda$. This unfavorable tendency in the calculation of 
$\lambda$ seems to prevail even in CaC$_6$~\cite{Mauri}.

The anisotropy of the critical magnetic field $H_{c2}$ was also a matter of 
interest, drawing attention of both experimentalists~\cite{Koike80,Iye,Roth85,
Dresselhaus89,Chaiken90} and theorists~\cite{Jishi91,Jishi92}. Note that the 
gap function $\Delta_{\bm p}$ defined in Eq.~(\ref{eq:F3}) has nothing to do 
with the anisotropic behavior of $H_{c2}$, though in developing a 
phenomenological theory~\cite{Jishi91,Jishi92}, some critical comments were 
made on the results of $\Delta_{\bm p}$~\cite{Takada82} with the assumption 
that the anisotropy in $H_{c2}$ should reflect on $\Delta_{\bm p}$. 

In search of higher $T_c$, many attempts have been made to synthesize new 
GIC superconductors such as NaC$_2$ ($T_c=5$K)~\cite{Belash87}, 
LiC$_2$ ($T_c=1.9$K)~\cite{Belash89b}, and alkali-metal amalgams like 
KHgC$_4$ ($T_c=0.73$K) and KHgC$_8$ ($T_c=1.90$K).~\cite{Alexander80,Tanuma81,
Koike81,Pendrys81,Alexander81,Iye}, but a larger enhancement of $T_c$ was not 
achieved until CaC$_6$ was found in 2005 with $T_c=11.5$K~\cite{CaC6}. 
Subsequently, many works have been done on alkaline-earth GIC 
superconductors~\cite{Emery,Kim1,Kim2,Lamura06,Hinks,Kurter,Kadowaki07,Sugawara09,
Valla09}, but no one has ever succeeded in synthesizing a new GIC with $T_c$ 
larger than 15.4K which was observed in CaC$_6$ under pressures~\cite{Takagi}. 
Thus some new idea seems to be needed to further enhance $T_c$. The present 
author hopes that the suggestions given in Sec.~8 help experimentalists 
synthesize a new GIC superconductor with $T_c$ much higher than 10K. 

\section{Standard Model for Superconductivity in GICs}
\label{sec:6}

\subsection{Characteristic Features of the System}

Basically because GICs are not recognized as strongly-correlated systems, 
the usual {\it ab initio} self-consistent band-structure calculation is 
very useful in elucidating the important features of the electronic 
structures of GICs in the normal state. According to such calculations, 
it is found that there is no essential qualitative difference between 
alkali and alkaline-earth GICs (see Fig.~\ref{fig-3}). The main common 
features among these GICs may be summarized in the following way: 
\begin{figure}[htbp]
\begin{center}
\includegraphics[width=8.5cm]{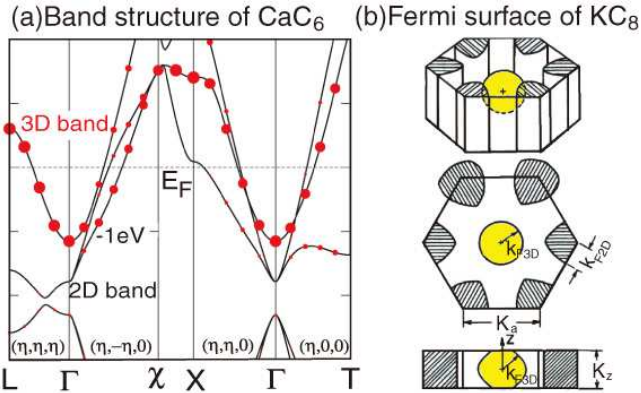}
\end{center}
\caption{(a)Band structure of CaC$_6$~\cite{Mauri}. (b)Fermi surface of 
KC$_8$~\cite{Wang91}. Both materials are characterized by the common feature 
that the electronic system is composed of the 2D $\pi$ bands of graphite and 
the 3D interlayer band. }
\label{fig-3}
\end{figure}
\begin{enumerate}
\item[a)]
In $M$C$_x$, each intercalant metal atom acts as a donor and changes from a 
neutral atom $M$ to an ion $M^{Z+}$ with valence $Z$. 
\item[b)]
The valence electrons released from $M$ will transfer either to the graphite 
$\pi$ bands or the three-dimensional (3D) band composed of the intercalant 
orbitals and the graphite interlayer states~\cite{Posternak84,Holzwarth84,
Koma86}. We shall define the factor $f$ as the branching ratio between these 
two kinds of bands. Namely, $Zf$ and $Z(1-f)$ electrons will go to the $\pi$ 
and the 3D bands, respectively. 
\item[c)]
The electrons in the graphite $\pi$ bands are characterized by the 
two-dimensional (2D) motion with a linear dispersion relation (known as a Dirac 
cone in the case of graphene) on the graphite layer. 
\item[d)]
The dispersion relation of the graphite interlayer band is very similar to 
that of the 3D free-electron gas, folded into the Brillouin zone of the 
graphite~\cite{Csanyi}. Thus its energy level is very high above the Fermi 
level in the graphite, because the amplitude of the wavefunction for this band 
is small on the carbon atoms. In $M$C$_x$, on the other hand, the cation 
$M^{Z+}$ is located in the interlayer position where the amplitude of the 
wavefunctions is large, lowering the energy level of the interlayer band 
below the Fermi level. The dispersion of the interlayer band is modified 
from that of the free-electron gas because of the hybridization with the 
orbitals associated with $M$, but generally it is well approximated by 
$\varepsilon_{\bm p}={\bm p}^2/2m^*-E_{\rm F}$ with an appropriate choice of 
the effective band mass $m^*$ and the Fermi energy $E_{\rm F}$. Here the value 
of $m^*$ depends on $M$; in alkali GICs, the hybridization occurs with 
s-orbitals, allowing us to consider that $m^*=m_e$, while in alkaline-earth 
GICs, the hybridization with d-orbitals contributes much, leading to 
$m^* \approx 3m_e$ in both CaC$_6$ and YbC$_6$, as revealed by 
the band-structure calculation~\cite{Mazin,Mauri}. 
\item[e)]
The value of $f$, which determins the branching ratio $Zf:Z(1-f)$, can be 
obtained by the self-consistent band-structure calculation. In KC$_8$, for 
example, it is known that $f$ is around $0.6$~\cite{Ohno79}. On the other hand, 
$f$ is about $0.16$~\cite{Mauri} in CaC$_6$, making the electron density 
in the 3D band $n$ increase very much. This increase in $n$ is 
easily understood by the fact that the energy level of the interlayer band 
is much lower with Ca$^{2+}$ than with K$^+$. The concrete numbers for 
$n$ are $3.5 \times 10^{21}$cm$^{-3}$ and $2.4 \times 10^{22}$cm$^{-3}$ 
for KC$_8$ and CaC$_6$, respectively, in which the difference in both $d$ and 
$x$ is also taking into account. 
\item[f)]
As inferred from experiments~\cite{PhysToday2,Takada82,Csanyi} and also from 
the comparison of $T_c$ calculated for each band~\cite{Takada82}, 
it has been concluded that only the 3D interlayer band is responsible for 
superconductivity. Note that LiC$_6$ does not exhibit superconductivity 
because no carriers are present in the 3D interlayer band, although the 
properties of LiC$_6$ are generally very similar to those of other 
superconducting GICs in the normal state. 
\end{enumerate}

\subsection{Microscopic Model for Superconductivity}

\begin{figure}[htbp]
\begin{center}
\includegraphics[width=8.0cm]{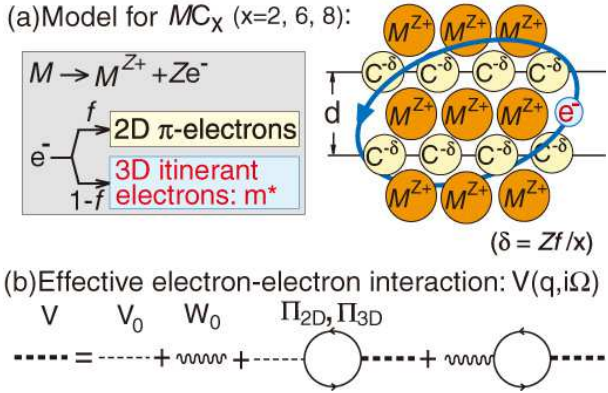}
\end{center}
\caption{(a)Simplified model to represent $M$C$_x$ superconductors. 
We consider the attraction between the 3D electrons in the interlayer band 
induced by polar-coupled charge fluctuations of the cation $M^{Z+}$ and 
the anion C$^{-\delta}$. (b)Diagrammatic representation of the equation 
in the RPA to calculate the effective electron-electron interaction 
$V({\bm p\!-\!p'},i\Omega)$, which will be substituted in Eq.~(\ref{eq:F5}) 
to evaluate the kernel of the gap equation.}
\label{fig-4}
\end{figure}

With these common features in mind, we can think of a simple model for 
the GIC superconductors, which is schematically shown in 
Fig.~\ref{fig-4}(a). Actually, exactly the same model was proposed in 
as early as 1982 by the present author for describing superconductivity in 
alkali GICs~\cite{Takada82}. 

In order to give some idea about the mechanism to induce an attraction between 
3D electrons in this model, let us imagine how each conducting 3D electron sees 
the charge distribution of the system. First of all, there are positively charged 
metallic ions $M^{Z+}$ with its density $n_M$, given by $n_M=4/3\sqrt{3}\,
a^2dx$, where $a$ is the bond length between C atoms on the graphite 
layer (which is 1.419\AA). Note that with use of this $n_M$, the density of the 
3D electrons $n$ is given by $(1-f)Zn_M$. There are also negatively charged 
carbon ions C$^{-\delta}$ with the average charge of $\delta\equiv -fZe/x$. 
Therefore the 3D electrons will feel a large electric field of the polarization 
wave coming from oscillations of $M^{Z+}$ and C$^{-\delta}$ ions created by 
either out-of-phase optic or in-phase acoustic phonons. 

We shall consider the coupling of those phonons with the 3D electrons in terms 
of the point-charge model, allowing us to write the phonon-exchange 
polar-coupled interaction $W_0({\bm q},\omega)$ for the scattering of the 3D 
electrons with momentum- and energy-transfers of ${\bm q}$ and $\omega$ as 
\begin{align}
 \label{eq:F8}
W_0({\bm q},\omega) =& V_0({\bm q})
\frac{\omega_p^2 (1\!-\!f)^2}{\omega^2 \!-\! \omega_{\rm LA}({\bm q})^2}
\nonumber \\
&+ V_0({\bm q})
\frac{\bar{\omega}_p^2(\bar{M}/M_M\!+\!f\bar{M}/xM_{C})^2} 
{\omega^2 \!-\! \omega_{\rm LO}({\bm q})^2},
\end{align}
with $\omega_p$ and $\bar{\omega}_p$ defined, respectively, as
\begin{eqnarray}
 \label{eq:F9}
\omega_p = \sqrt{{4\pi e^2 Z^2 n_M \over M_M+xM_C}}\quad {\rm and}\ 
\bar{\omega}_p = \sqrt{{4\pi e^2 Z^2 n_M \over \bar{M}}},
\end{eqnarray}
where $M_M$ and $M_C$ are, respectively, the atomic masses of $M$ and C, 
$\bar{M}\,(=M_MxM_C/(M_M+xM_C)$) is the reduced mass of $M$C$_x$, 
$\omega_{\rm LO}({\bm q})$ 
and $\omega_{\rm LA}({\bm q})$ are the energies of LO- and LA-phonons, 
respectively, and $V_0({\bm q})$ is the bare Coulomb interaction 
$4\pi e^2/{\bm q}^2$. (The subscript $0$ indicates that it is the 
bare interaction to be screened by both 2D and 3D mobile electrons.) 

Owing to the coupling with valence electrons, both 
$\omega_{\rm LO}({\bm q})$ and $\omega_{\rm LA}({\bm q})$ depends on $f$, 
but the $f$-dependence is not important, if we write 
the phonon-mediated interaction in terms of the corresponding transverse 
phonon energies, $\omega_{\rm TO}({\bm q})$ and $\omega_{\rm TA}({\bm q})$. 
Thus we specify the phonon energies in terms of $\omega_{\rm TO}({\bm q})$ 
and $\omega_{\rm TA}({\bm q})$. In actual calculations, we assume that 
$\omega_{\rm TO}({\bm q}) = \omega_t (=$ constant) and 
$\omega_{\rm TA}({\bm q}) = c_t |{\bm q}|$ with $\omega_t$ of the order 
of 150K and $c_t$ of the order of $10^5$cm\,s$^{-1}$ for the oscillation 
perpendicular to the graphite plane. 

\subsection{Calculation of $T_c$ for Alkali-Doped GICs}

\begin{figure}[htbp]
\begin{center}
\includegraphics[width=5.5cm]{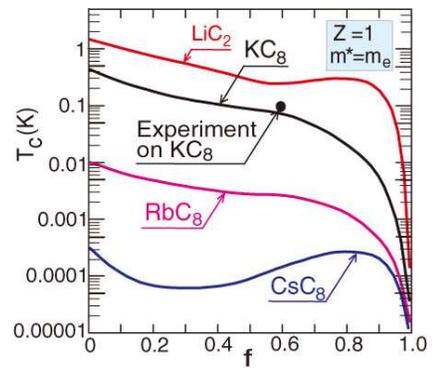}
\end{center}
\caption{Calculated results for $T_c$ as a function of the branching ratio $f$ 
for alkali GIC superconductors in which $Z=1$ and $m^*=m_e$.}
\label{fig-5}
\end{figure}

By combining this polar-phonon-mediated attractive interaction 
$W_0({\bm q},\omega)$ with the bare Coulomb interaction between electrons 
$V_0({\bm q})$ on the same footing and considering the polarization effects 
of both 2D and 3D electrons, we faithfully calculate $V({\bm q},\omega)$ the 
effective interaction between 3D electrons in the RPA (see, Fig.~\ref{fig-4}(b)). 
The obtained $V({\bm q},\omega)$ is put into the kernel, Eq.~(\ref{eq:F5}), 
of the gap equation~(\ref{eq:F3}) to obtain $T_c$ from first principles. 
The calculated results for $T_c$ in alkali GICs are plotted as a function 
of $f$ in Fig.~\ref{fig-5} to find that the overall magnitude of $T_c$ is in the 
range of $0.1-0.01$K for $f \approx 0.5$, in good agreement with experiment. 
Note that smaller values of $T_c$ are obtained for heavier alkali atoms 
because of the smaller couplings as characterized by both $\omega_p$ and 
$\bar{\omega}_p$. This success indicates that the present simple model applies 
well at least to alkali GIC superconductors. 

\section{Superconductivity in Alkaline-Earth GICs}
\label{sec:7}

\subsection{CaC$_6$}
\label{sec:7_1}

Now let us consider alkaline-earth GIC superconductors. We shall investigate 
them by adopting the same simple model with using exactly the same calculation 
code developed in 1982 in order to see whether the model and therefore the 
piture on the mechanism of superconductivity successfully applied to alkali 
GIC superconductors can also be relevant to these newly-synthesized 
superconductors or not~\cite{Takada09a,Takada09b}. The parameters specifying 
the model will be changed in the following way, if CaC$_6$ is considered 
instead of KC$_8$:
\begin{enumerate}
\item[a)]
Because the valence $Z$ changes from monvalence to divalence, the atractive 
interaction $W_0$, which is in proportion to $Z^2$, increases by four times. 
\item[b)]
The interlayer distance $d$ decreases from $5.42$\AA\ to $4.524$\AA, 
so that the 3D electron density $n$ increases. 
\item[c)]
The factor $f$ to determine the branching ratio decreases from about $0.6$ 
to $0.16$. 
\item[d)]
The effective band mass for the 3D interlayer band $m^*$ increases from 
$m_e$ to about $3m_e$. 
\item[e)]
The atomic number of the ion $A$ hardly changes from $39.1$ to $40.1$. 
\end{enumerate}
\begin{figure}[htbp]
\begin{center}
\includegraphics[width=5.5cm]{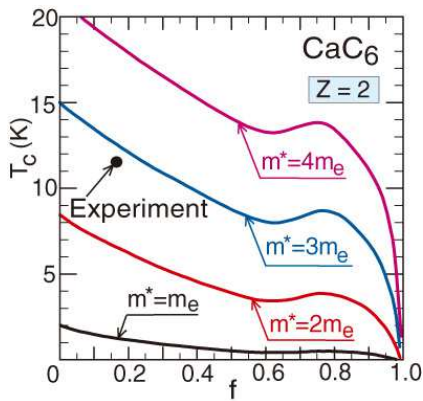}
\end{center}
\caption{Calculated $T_c$ as a function of $f$ for $m^*$ in the 
range of $m_e-4m_e$ with other parameters suitably chosen for CaC$_6$. The 
experimental result is reproduced well, if we choose $m^* \approx 3m_e$.}
\label{fig-6}
\end{figure}

With paying attention to these changes of the parameters, we have calculated 
$T_c$ for CaC$_6$ as a function of $f$. The results are plotted in 
Fig.~\ref{fig-6}, from which we can learn the following points: 
\begin{enumerate}
\item[1)]
Overall, $T_c$ becomes higher for smaller $f$. This can be understood by the 
fact that the screening effect due to the 2D $\pi$ electrons, which makes the 
polar-coupled interaction weak, becomes smaller with the decrease of $f$. 
\item[2)]
The enhancement of $T_c$ by about one order is brought about by doubling $Z$, 
if $m^*$ is kept to be the same value. 
\item[3)]
The enhancement of $T_c$ by about one order is also brought about by tripling 
$m^*$ from $m_e$ to $3m_e$, if $Z$ is taken as $Z=2$. 
\end{enumerate}

Based on these observations, we can conclude that the enhancement of $T_c$ 
in CaC$_6$ by about a hundred times from that in KC$_8$ is brought about by 
the combined effects of doubling $Z$ and tripling $m^*$. In this respect, 
the value of $m^*$ is very important. Appropriateness of $m^* \approx 3m_e$ 
is confirmed not only from the band-structure calculations~\cite{Mazin,Mauri} 
but also from the measurement of the electronic specfic heat~\cite{Kim1} 
compared with the corresponding one for KC$_8$~\cite{Mizutani78}.

\subsection{Other alkaline-earth GICs}
\label{sec:7_2}

\begin{figure}[htbp]
\begin{center}
\includegraphics[width=5.5cm]{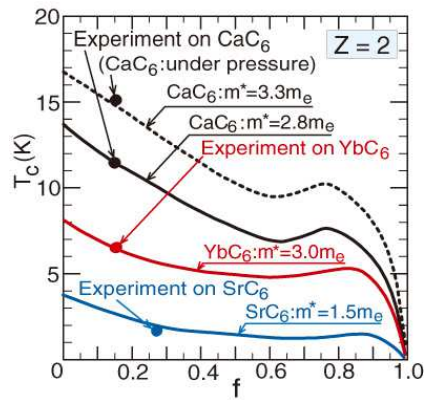}
\end{center}
\caption{Calculated $T_c$ as a function of $f$ for alkaline-earth 
GICs with $m^*$ determined so as to reproduce $E_{\rm F}$ provided by the 
band-structure calculation.}
\label{fig-7}
\end{figure}

Similar calculations are done for other alkaline-earth GIC superconductors 
as shown in Fig.~\ref{fig-7} in which $m^*$ is determined so as to reproduce 
$E_{\rm F}$ supplied by the band-structure calculation. We see that although 
we give $T_c$ a little larger than the experimental one for SrC$_6$, overall 
good agreement is obtained between theory and experiment, implying that our 
simple model may be regarded as the standard one for describing the mechanism 
of superconductivity in GICs. 

Here a note will be added to the case of YbC$_6$; the basic parameters such 
as $Z$, $f$, and $m^*$ for YbC$_6$ are about the same as those for CaC$_6$, 
according to the band-structure calculation. The only big change can be seen 
in the atomic mass; Yb (in which $A=173.0$) is much heavier than that of Ca 
by about four times, indicating weaker couplings between electrons and polar 
phonons as just in the case of comparison between KC$_8$ and CsC$_8$. In fact, 
$T_c$ for YbC$_6$ becomes about one half of the corresponding result for 
CaC$_6$, which agrees well with experiment. One way to understand this 
difference is to regard it as an isotope effect with $\alpha 
\approx 0.5$~\cite{Mazin}. 

\subsection{BaC$_6$}
\label{sec:7_3}

The experimental results for $T_c$ in the alkaline-earth GICs treated 
in Fig.~\ref{fig-7} are also well reproduced by the the conventional 
Eliashberg theory in which the McMillan's formula for $T_c$ is employed with 
use of the electron-phonon coupling constant $\lambda$, the average phonon 
energy $\omega_0$, and the Coulomb pseudopotential $\mu^*$ with 
its conventional value of $\mu^*=0.14$. The two parameters, $\lambda$ and 
$\omega_0$, are determined by the first-principles calculation 
of the Eliashberg function $\alpha^2F(\omega)$~\cite{Calandra06}. 

This success of the Eliashberg theory is, however, limited; the same theory 
predicts that BaC$_6$ superconducts at $T_c=0.23$K, but it turns out that 
superconductivity does not appear at least down to 80mK~\cite{Nakamae08}. 
In search of the reason for this discrepancy between theory and experiment, 
the phonon structure is extensively studied in comparison with the 
case of CaC$_6$~\cite{Astuto10,Walters11}, but no persuasive reason has 
been found. In the present author's view, this failure is directly connected 
with the problem of obtaining an unrealistically large contribution from the 
intralayer high-energy carbon oscillations to $\alpha^2F(\omega)$, as 
mentioned in the first paragraph in Sec.~\ref{sec:5}. In fact, $\omega_0
\!=\!22.44$meV is obtained for BaC$_6$~\cite{Calandra06}, which 
is much higher than the energy of Ba oscillations ($\approx 8$meV), 
indicating that the carbon modes are responsible for the unsuccessful 
prediction of $T_c=0.23$K in the conventional Eliashberg theory. 

\begin{figure}[htbp]
\begin{center}
\includegraphics[width=8.4cm]{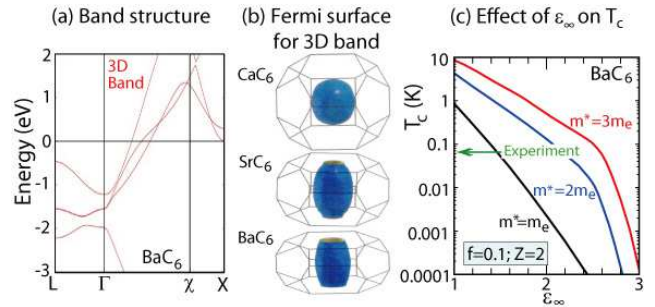}
\end{center}
\caption{(a) Band structure of BaC$_6$. (b) The Fermi surfaces for the 3D 
interlayer bands in CaC$_6$, SrC$_6$, and BaC$_6$~\cite{Calandra06}. 
(c) Calculated $T_c$ as a function of the optic dielectric constant 
$\varepsilon_{\infty}$ for BaC$_6$ with $m^*$ in the range $m_e-3m_e$. 
Note that $T_c$ is 65mK experimentally~\cite{Heguri}.}
\label{fig-8}
\end{figure}

Very recently BaC$_6$ is discovered to exhibit superconductivity with 
$T_c=65$mK~\cite{Heguri}. In the framework of the standard model, $m^*$, 
$f$, and $Z$ are the important parameters to be determined by the 
band-structure calculation, from which we see that we may take $Z=2$ and 
$f$ in the range of $0.1-0.3$, the same situation as those in CaC$_6$ and 
SrC$_6$. As for $m^*$, it becomes smaller than $3m_e$, because the 
interlayer 3D band of graphite is hybidized with the more itinerant 5d 
orbitals in BaC$_6$ compared with the 3d ones in CaC$_6$; if we 
compare the dispersion relation for the 3D band along $\Gamma\chi$ direction 
for BaC$_6$ as shown in Fig.~\ref{fig-8}(a) with that for CaC$_6$ given 
in Fig.~\ref{fig-3}(a), we find that $m^* \approx 1.9m_e$. In addition, 
the 3D band at $L$ point is located below the Fermi level due to the 
shorter Brilloiun zone (or equivalently the longer lattice constant) 
for BaC$_6$, indicating that some portion of the otherwise spherical Fermi 
surface is truncated or missing, as seen in Fig.~\ref{fig-8}(b) which 
displays the Fermi surfaces for the 3D interlayer bands in CaC$_6$, SrC$_6$, 
and BaC$_6$. Because of this truncation or missing, the virtual multiple 
scatterings to form the Cooper pairs are restricted, so that $T_c$ will be 
suppressed from the value obtained in the standard model. Note that, though 
its size is much smaller, this truncation or missing is also seen in SrC$_6$ 
and thus the reduction of $T_c$ in experiment for SrC$_6$ from that predicted 
in the standard model may be ascribed to this effect. 

This missing of the sperical Fermi surface implies the reduction of the density 
of states for the 3D interlayer band at the Fermi level and therefore it might 
be effectively taken into account by the reduction of $m^*$ from that in the 
band-structure calculation. Probably $m^* \approx 1.5m_e$ will be a reasonable 
choice. With this idea in mind, we have calculated $T_c$ for BaC$_6$ with $m^*$ 
in the range $m_e-2m_e$ to find that, irrespective of $f$ taken in the range of 
$0.1-0.5$, the obtained $T_c$ is always larger than 0.1K, which is about the 
same as that in the Eliashberg theory but is much higher than the experimental 
value. Thus we need to look for another crucial parameter in the standard model 
to explain the experimental value of $T_c$. 

Basically, the standard model assumes the polar-coupling phonon mechanism of 
superconductivity in which, in general, the effect of the optic dielectric 
constant $\varepsilon_{\infty}$ should be included in the theory and can be 
treated by changing $e^2$ into $e^2/\varepsilon_{\infty}$ in Eqs.~(\ref{eq:F8}) 
and (\ref{eq:F9}). Physically $\varepsilon_{\infty}$ is determined by the 
magnitude of core polarization of constituent atoms or ions. For light atoms 
like carbon, the core polarization is negligibly small and thus we may well take 
$\varepsilon_{\infty}$ as unity. Even for Ca$^{2+}$, its polarizability is 
about 3.2 in atomic units~\cite{Mitroy10}, leading to $\varepsilon_{\infty}
=1.07$. For heavy atoms, however, it can never be neglected; for Ba, the 
polarizabilities are, respectively, 124 and 10.5 for Ba$^+$ and Ba$^{2+}$, which 
correspond, respectively, to $\varepsilon_{\infty}=3.8$ and 1.24. By combining 
these numbers for $\varepsilon_{\infty}$ with the fact that the 3D interlayer band 
is completely occupied in the $\Gamma-L$ direction, making some portion of the 
released 6s electrons actually localize near the Ba$^{2+}$ site, leading 
effectively to the state of Ba$^{(2-\delta)+}$, we may assume that the effective 
value for $\varepsilon_{\infty}$ is in the range of $1.5-2.0$. Then, as we can 
see in Fig.~\ref{fig-8}(c), $T_c$ obtained in the standard model with $m^*\approx 
1.5m_e$ fits well with the experimental one.

\section{Prediction of the Optimum $T_c$ in GICs}
\label{sec:8}

As we have seen so far, our standard model could have predicted $T_c =11.5$K 
for CaC$_6$ in 1982 and it is judged that its predictive power is very high. 
Incidentally, the author did not perform the calculation of $T_c$ for 
CaC$_6$ at that time, partly because he did not know a possibility to synthesize 
such GICs, but mostly because the calculation cost was extremely high in 
those days; a rough estimate shows that there is acceleration in computers by 
at least a millon times in the past three decades. This huge improvement 
on computational environments is surely a boost to making such first-principles 
calculations of $T_c$ as reviewed in Secs.~3 and 4. 

\begin{figure}[htbp]
\begin{center}
\includegraphics[width=7.5cm]{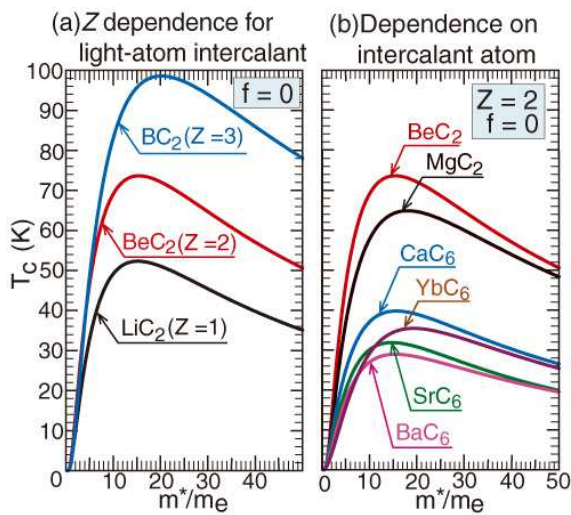}
\end{center}
\caption{Prediction of $T_c$ as a function of $m^*$ for various 
GICs in pursuit of optimum $T_c$. We assume the fractional factor $f=0$.}
\label{fig-9}
\end{figure}

In any way, encouraged by this success in reproducing $T_c$ in alkaline-earth 
GICs, we have explored the optimum $T_c$ in the whole family of GICs by widely 
changing various parameters involved in the microscopic Hamiltonian. Examples 
of the calculated results of $T_c$ are shown in Fig.~\ref{fig-9}(a) and (b), 
in which $f$ is fixed to zero, the optimum condition to raise $T_c$, and $d$ is 
tentatively taken as 4.0\AA. From this exploration, we find that the most 
important parameter to enhance $T_c$ is $m^*$. In particular, we need $m^*$ 
larger than at least $2m_e$ to obtain $T_c$ over 10K, irrespective of any 
choice of other parameters, and $T_c$ is optimized at $m^*$ near $15m_e$. The 
optimized $T_c$ depends rather strongly on the parameters to control the 
polar-coupling strength such as $Z$ and the atomic mass $A$; if we choose a 
trivalent light atom such as boron to make $\omega_t$ large, the optimum $T_c$ 
is about 100K, but the problem about the light atoms is that $m^*$ will never 
become heavy due to the absence of either d or f electrons. Therefore we do 
not expect that $T_c$ would become much larger than 10K, even if BeC$_2$ or 
BC$_2$ were synthesized. From this perspective, it will be much better to 
intercalate Ti or V, rather than Be or B. Taking all these points into account, 
we suggest synthesizing three-element GICs providing a heavy 3D electron system 
by the introduction of heavy atoms into a light-atom polar-crystal environment.

\section{Conclusion}
\label{sec:9}

In this chapter, by taking account of the common features elucidated by both 
the band-structure calculation and the various measurements on the normal-state 
properties, we have constructed the standard model pertinent for the description 
of the mechanism of superconducitity in metal GICs and then 
made first-principles calculations of $T_c$ in the $G_0W_0$ scheme, directly 
from the microscopic Hamiltonian representing the standard model. With suitably 
choosing the parameters in the microscopic Hamiltonian, we have found 
surprisingly good agreement between theory and experiment for both alkali and 
alkaline-earth GICs, in spite of the fact that $T_c$ varies more than three 
orders of magnitude. In this way, we have clarified that superconductivity 
in metal GICs can be understood by the picture that the 3D electrons in the 
interlayer band supplied by the ionization of metals feel the attractive 
interaction induced by the virtual exchange of the polar-coupled phonons of 
the metal ions. We have also predicted a further enhancement of $T_c$ well 
beyond 10 K with giving some suggestions to realize such superconductors 
in the family of GICs. 

By first-principles we usually mean the calculations based on not the model 
but the first-principles Hamiltonian. Thus it might be considered as 
inappropriate to call the present $G_0W_0$ scheme first-principles, but it is 
not an easy task to specify the key parameters to control $T_c$ by just 
implementing the calculations based on the first-principles Hamiltonian. We 
can identify the importance of the parameters, $m^*$ and $Z$, only through the 
calculations based on the model Hamiltonian, leading to the better and 
unambiguous understanding of the mechanism of superconductivity without 
involved too much into the very details of each system which sometimes 
obscure the essence in first-principles approaches. Besides, because of the 
errors involved in the numerical calculations of normal-state 
properties as mentioned in Sec.~4, more accurate results of $T_c$ will be 
obtained by way of a suitable model Hamiltonian rather than directly from 
the first-principles one. 

As a project in the future, it would be important to construct a more powerful 
scheme for the first-principles calculation of $T_c$ by the combination of 
the schemes in Secs.~3 and 4, based on which we may make more detailed 
suggestions to synthesize GIC superconductors with $T_c$ much larger than 10K. 



\end{document}